\documentclass{article}

 \usepackage[preprint]{neurips_2026}

\usepackage[utf8]{inputenc} 
\usepackage[T1]{fontenc}    
\usepackage[hidelinks]{hyperref} 
\usepackage{url}            
\usepackage{booktabs}       
\usepackage{amsmath}        
\usepackage{amsfonts}       
\usepackage{nicefrac}       
\usepackage{microtype}      
\usepackage[table]{xcolor}  
\usepackage{xspace}         
\usepackage{graphicx}       
\usepackage{wrapfig}        
\usepackage{tabularx}       
\usepackage{caption}        
\usepackage{needspace}      
\usepackage{colortbl}       
\usepackage{placeins}       
\usepackage{float}          
\usepackage{etoc}           

\definecolor{tablerow}{gray}{0.93}

\renewcommand{\arraystretch}{0.9}

\newcommand{\ourbench}{SiliconBench\xspace}

\title{SiliconBench: Speed, Memory, and Fidelity for LLM Serving on Unified-Memory Desktops}

\author{%
  \textbf{Ranran Haoran Zhang}\textsuperscript{1}\quad
  \textbf{Aysa Xuemo Fan}\textsuperscript{2}\quad
  \textbf{David Munhá Correia}\textsuperscript{3}\\[0.3em]
  \textbf{Alex Cheema}\textsuperscript{3}\quad
  \textbf{Rui Zhang}\textsuperscript{1}\\[0.6em]
  {\normalfont\small \textsuperscript{1}Penn State University\quad
  \textsuperscript{2}University of Illinois Urbana-Champaign\quad
  \textsuperscript{3}EXO Labs}\\[0.3em]
  {\normalfont\small Email:
  \href{mailto:hzz5361@psu.edu}{\nolinkurl{hzz5361@psu.edu}}\quad
  \href{mailto:rmz5227@psu.edu}{\nolinkurl{rmz5227@psu.edu}}}\\[0.3em]
  {\normalfont\small Website: \url{https://ranranhaoranzhang.com/siliconbench/}}\\
  {\normalfont\small Code: \url{https://github.com/WindChimeRan/SiliconBench}}
}

\hypersetup{
  pdftitle={SiliconBench: Speed, Memory, and Fidelity for LLM Serving on Unified-Memory Desktops},
  pdfauthor={Ranran Haoran Zhang, Aysa Xuemo Fan, David Munhá Correia, Alex Cheema, Rui Zhang}
}

\begin{document}
\etocdepthtag{main}

\maketitle

\begin{abstract}

  Concurrent local LLM serving on unified-memory desktops must preserve
  memory headroom and output fidelity, which speed-only rankings overlook.
  We introduce \ourbench, which evaluates nine Apple Silicon serving
  engines through three lenses: speed, memory, and fidelity.
  We evaluate chat and agent serving on Qwen3, Qwen3.5, and Gemma~4.
  We use a classification task to check for quality regressions against
  an NVIDIA reference. DGX Spark provides a complementary
  serving-performance reference. Three desiderata guide
  interpretation: serving architecture readiness, memory discipline, and
  multi-node scaling.
  On Qwen3-0.6B, \texttt{vllm-metal} alone more than doubles throughput
  on both workloads from concurrency 1 to 16. CUDA vLLM and SGLang
  show stronger concurrency scaling on the same prompts.
  Explicit memory budgets do not guarantee memory headroom: two stacks
  complete every request while memory use approaches physical capacity
  and throughput declines. The newer model architectures have
  narrower engine support. Their evaluated implementations match the
  fidelity reference. Only three stacks satisfy the completion,
  fidelity, and model-coverage gates. Comparisons on larger dense and MoE
  models reinforce the importance of scheduling prompt processing alongside
  ongoing generation: \texttt{vllm-metal}'s packed prefill--decode path
  maintains lower first-token latency than \texttt{omlx} under concurrent
  load. In the tested two-machine
  configurations, tensor parallelism over Thunderbolt RDMA scales
  while pipeline parallelism over TCP regresses.
  We release benchmark code, per-run results, and maintenance journals,
  supported by a workflow combining bounded agent fixes with human review.
\end{abstract}

\section{Introduction}
\label{sec:intro}

Unified-memory desktops, including Apple Silicon systems and NVIDIA
DGX Spark, are becoming platforms for local LLM serving. They use
low-power DDR (LPDDR) memory shared by CPU and GPU, whereas many
datacenter accelerators use dedicated high-bandwidth memory
(HBM).\footnote{Hardware examples: \href{https://www.ifixit.com/News/54122/macbook-pro-2021-teardown}{Apple M1 Pro (LPDDR5)},
\href{https://docs.nvidia.com/dgx/dgx-spark/hardware.html}{DGX Spark (LPDDR5x)}, and
\href{https://docs.nvidia.com/enterprise-reference-architectures/hgx-ai-factory-h100-h200-b200/latest/components.html}{NVIDIA H100/H200/B200 (HBM)}.}
Unified-memory consumer hardware now reaches 512\,GB, capacities historically
confined to data-center accelerators.
Within this broader class, our main audit focuses on Apple Silicon, where
local LLM servers share this pool with macOS and foreground applications.
The scale of the Mac ecosystem for local LLMs motivates this focus:
hundreds of thousands of annual Mac installs of Ollama,
llama.cpp, and LM Studio; 1.45M monthly PyPI downloads of
\texttt{mlx\_lm}, which targets Apple Silicon exclusively; and
4{,}710 community-converted models on Hugging Face's mlx-community
hub.\footnote{Ecosystem statistics collected on 2026-05-03 from
Homebrew Analytics (\url{formulae.brew.sh}), PyPI Stats
(\url{pypistats.org}), and the mlx-community organization page on
Hugging Face.}

At least nine actively-developed inference stacks compete on this
hardware (Table~\ref{tab:features}). Users choose among them primarily
by throughput, but speed alone does not show how much memory remains for
everyday use. On Apple Silicon, model weights and inference caches share
RAM with the user's browser, IDE, and OS. An engine can therefore rank
highly in throughput while leaving little room for these applications.
As their combined memory demand approaches the machine's capacity,
memory pressure can slow inference and foreground applications or cause
requests to fail. Concurrency adds a second concern: single-request speed
does not establish how an engine performs under parallel load. A single
coding agent can issue 4 to 16 simultaneous requests to a local server,
increasing scheduling and memory demands. Stacks that lack continuous
batching or memory discipline may slow down or crash under this load.
Even when requests complete, an engine can produce lower task-level scores
than a reference implementation running the same model. No existing
benchmark evaluates these stacks jointly on speed, memory, and output fidelity.

We introduce \ourbench to evaluate concurrent chat and agent serving
across nine Apple Silicon engines. It measures throughput, latency,
memory use, and request completion, with task-level fidelity assessed
separately against an NVIDIA A100 reference. Qwen3-0.6B provides a common
baseline; Qwen3.5-0.8B and Gemma-4-E4B-it test support for newer architectures
(\S\ref{sec:results_fidelity}, Appendix~\ref{sec:appendix_speed_extra}).
A complementary DGX Spark track evaluates shared engine families
on the same models and workloads
(\S\ref{sec:results_dgxspark}, Appendix~\ref{sec:appendix_dgxspark}).
Additional studies examine larger dense/MoE models and multi-node
serving (\S\ref{sec:results_large_models}, \S\ref{sec:results_d3},
Appendix~\ref{sec:appendix_large_models}).

\begin{table}[t]
\centering
\small
\setlength{\tabcolsep}{4pt}
\renewcommand{\arraystretch}{1.12}
\rowcolors{2}{tablerow}{white}
\begin{tabularx}{\linewidth}{@{}>{\columncolor{white}[0pt][\tabcolsep]}l>{\raggedright\arraybackslash}p{0.90in}>{\raggedright\arraybackslash}X>{\raggedright\arraybackslash}p{0.92in}>{\columncolor{white}[\tabcolsep][0pt]\raggedright\arraybackslash}p{0.72in}@{}}
\toprule
\rowcolor{white}
\textbf{Engine} & \textbf{Prompt batching} & \textbf{KV-cache layout} & \textbf{Prefill / decode} & \textbf{Memory policy} \\
\midrule
\texttt{llama.cpp} & Packed tokens & Preallocated KV cells & Same pass & Advisory hint \\
\texttt{ollama} & Packed tokens & Preallocated KV cells & Same pass & Advisory hint \\
\texttt{mlx\_lm} & Padded batch & Contiguous per batch & Separate passes & Unbounded \\
\texttt{vllm-metal} & Packed tokens & Paged & Same pass & Explicit cap \\
\texttt{vllm-mlx} & Padded batch & Contiguous per batch & Separate passes & Explicit cap \\
\texttt{omlx} & One prompt & Contiguous per batch & Separate passes & Explicit cap \\
\texttt{sglang} & One prompt & Contiguous per request & Separate passes & Explicit cap \\
\texttt{hf\_transformers} & Packed tokens & Paged & Same pass & Advisory hint \\
\texttt{mistral.rs} & Padded batch & Contiguous per sequence & Separate passes & Explicit cap \\
\bottomrule
\end{tabularx}
\caption{\textbf{Serving engines differ in batching and memory management.}
Overview of the nine Apple Silicon configurations. Prompt batching
describes how prefills share a forward pass; packed batches omit
padding. Prefill/decode states whether a pass can combine both
phases. KV-cache layout covers active requests. Memory policies are
declared settings; enforcement is evaluated in
\S\ref{sec:results_memory}.}
\label{tab:features}
\end{table}

Dependency failures and a template error in five early runs show why
serving benchmarks need continued software validation. \ourbench combines
agent-proposed adapter fixes with human review before regenerating official
results (\S\ref{sec:maintenance_casestudy}).

\textbf{Chat-throughput rankings obscure differences in concurrent agent
serving.} On Qwen3-0.6B, \texttt{vllm-metal} alone more than
doubles throughput on both chat and agent prompts from concurrency
1 to 16. Only three Apple Silicon stacks satisfy the audit's completion,
fidelity, and model-coverage criteria.

We propose three desiderata for Apple Silicon serving: (D1) serving
architecture readiness, supporting new model architectures and serving
concurrent requests efficiently; (D2) memory discipline, using shared memory
efficiently while preserving headroom for macOS and foreground applications;
and (D3) multi-node scaling, distributing inference efficiently across
machines to serve models beyond one machine's memory capacity. We measure speed, memory, and
fidelity to assess current engines, and use these desiderata to interpret
their strengths and remaining gaps.

\needspace{6\baselineskip}
Our contributions:
\begin{itemize}
  \item \textbf{Three desiderata for Apple Silicon serving}:
    architecture readiness, memory discipline, and multi-node
    scaling, supported by benchmark measurements
    (\S\ref{sec:desiderata}).
  \item \textbf{A benchmark for joint evaluation} of concurrency scaling,
    memory use, and task-level fidelity across nine Apple Silicon stacks,
    with a complementary CUDA reference track
    (\S\ref{sec:baselines}).
  \item \textbf{A maintenance workflow for continued validation across
    software updates}, combining bounded agent-proposed adapter fixes
    with human review
    (\S\ref{sec:fidelity_pillar}).
\end{itemize}

\section{Related Work}
\label{sec:related_work}

\paragraph{LLM serving systems.}
Continuous batching~\citep{yu2022orca}, paged KV-cache management~\citep{kwon2023efficient}, chunked prefill with stall-free decode~\citep{agrawal2024sarathi}, and prefix-sharing via radix-tree indexing~\citep{zheng2024sglang} compose the serving stack now standard on CUDA. These techniques were developed mainly for CUDA datacenter systems, where dedicated VRAM and NVLink/NCCL are common. DGX Spark pairs CUDA with a unified CPU/GPU memory pool, providing a desktop reference for these serving capabilities. Distributed inference also spans consumer hardware: Petals~\citep{borzunov2023petals} targets collaborative GPU serving, while Prima.cpp~\citep{li2025prima} supports heterogeneous home clusters including Apple Metal devices. \ourbench audits concurrent serving across nine Apple Silicon stacks and evaluates three multi-node systems on a fixed workload.

\paragraph{Consumer and Apple Silicon inference benchmarks.}
Recent work benchmarks LLM inference on consumer hardware: comparing stacks on Apple Silicon~\citep{rajesh2025production,barrios2026native,benazir2025apple}, profiling CPU--GPU execution trade-offs~\citep{zhang2025challenging}, evaluating custom Metal kernels for long-context KV attention~\citep{vegasena2026opentq}, studying multi-node expert parallelism across Mac Studios~\citep{chen2024moe}, and adding cross-hardware breadth across accelerator families~\citep{stuhlmann2025bench360}. These efforts primarily measure throughput and task quality on single stacks or narrow hardware slices; \ourbench covers nine stacks on the audited platform, anchors them against their CUDA-native siblings on DGX Spark through three shared-engine bridge pairs, and evaluates speed, memory, and task-level fidelity.

\paragraph{Output drift and live benchmarks.}
LLM output drift arises from kernel nondeterminism, mixed-precision effects, and cross-provider divergence~\citep{he2025defeating,yuan2025give,khatchadourian2025llm,anthropic2025postmortem}. A separate line of work addresses benchmark obsolescence by refreshing prompts on a schedule~\citep{white2024livebench,jain2024livecodebench,zhang2025swe}. \ourbench keeps prompts fixed while updating framework adapters through the maintainer agent and reviewed community PRs; each published snapshot includes fidelity.

\section{Three Desiderata for LLM Serving on Apple Silicon}
\label{sec:desiderata}

We examine how current engines address these desiderata under two platform
constraints: memory shared with macOS and foreground applications, and the
absence of an intra-box multi-GPU path.

%
%

\subsection{D1: Serving architecture readiness}
\label{sec:attention_variants}

\emph{Serving new model architectures efficiently under concurrent load
requires compatible attention implementations and batching support.}

We first explain backend support for new models, then examine batching
under load and illustrate coverage gaps with recent attention architectures.

\paragraph{Backend support for new attention operations.}
CUDA benefits from a mature kernel-development ecosystem, including
CUDA C++, CUTLASS, and DSLs such as Triton, CuTe DSL, and
TileLang~\citep{triton,cutlassdsl,tilelang}. Metal code generation remains
less mature, increasing the implementation and tuning effort required to
support new attention architectures~\citep{tilelangmetal}. This tooling gap
helps explain why model support remains uneven across Apple Silicon
engines. We assess architecture readiness through model coverage and
performance under concurrent load.

\paragraph{Batching under load.}
Efficient concurrent serving depends on three implementation choices
made by every engine (Table~\ref{tab:features}).
\emph{Query layout} determines how requests' query tokens share a batch:
packed queries process $N_q=\sum_{b=1}^{B}q_b$ real token rows for $B$
requests with query lengths $q_b$, whereas padded batches process
$B T_{\max}$ rows, with $T_{\max}=\max_b q_b$. Some engines instead
prefill one prompt at a time.
\emph{Active-KV allocation} determines how active requests' caches are
stored: in paged blocks, contiguous storage, or preallocated KV cells.
Paging allows cache growth without reshaping contiguous storage; it does
not determine query layout.
\emph{Scheduler step composition} determines whether prefill and decode
share a forward pass: mixed steps combine both, whereas separate steps
process them in different passes.

These choices combine differently across engines. \texttt{omlx} avoids
prefill padding by processing one prompt at a time, uses contiguous
per-batch KV storage, and interleaves prefill chunks with batched decode
in separate forward passes. \texttt{vllm-metal} and
\texttt{hf\_transformers} combine packed queries, paged KV, and mixed
steps in the audited configurations. Their different measured scaling
shows that this combination alone does not determine performance
(\S\ref{sec:results_d1}).

\paragraph{New attention architectures.}
New model architectures impose requirements on both attention operators
and cache organization. Qwen3.5/3.6 combine recurrent attention state with
full-attention KV caches~\citep{gateddeltanet,qwen35}, while Gemma~4 shares
KV across layers~\citep{yoco}. Supporting these architectures requires
corresponding changes in the serving backend. As of April 2026, only
llama.cpp and vllm-metal supported both families; by our August
2026 campaign five of the nine stacks do
(Tables~\ref{tab:speed_qwen35_chat}
and~\ref{tab:speed_gemma4_chat}), while mistral.rs still lacks a
DeltaNet kernel and sglang's MLX backend serves neither family: it
has no hybrid-SSM kernels for Qwen3.5, and its text-only loader
rejects Gemma~4's multimodal checkpoint. Upstream sglang serves
both on CUDA (\S\ref{sec:results_dgxspark}), so the coverage gap
belongs to the Metal backend rather than the engine.

%
%
%
%
%
%
%
%

\subsection{D2: Memory discipline on unified-memory systems}
\label{sec:memory_management}

\emph{Inference engines must manage their total memory footprint to preserve
headroom for macOS and foreground applications.}

Model weights, live caches, temporary buffers, and retained allocator
buffers all contribute to this footprint.

Allocating too much KV memory risks system slowdowns or allocation
failures; too little limits serving capacity. Queueing new requests when
the KV pool is full can bound cache growth. We examine KV allocation
policies, then consider how allocator retention and padding can further
inflate the memory footprint.

\paragraph{Three allocation strategies in the wild.}
A \emph{profile-and-claim} policy designed for dedicated
VRAM~\citep{kwon2023efficient} needs adaptation when the available
pool changes with foreground use. The Memory policy column of
Table~\ref{tab:features} distinguishes explicit caps, advisory-hint sizing,
and unbounded growth. \texttt{vllm-metal}'s paged path
and \texttt{mistral.rs} expose explicit caps. \texttt{llama.cpp},
\texttt{ollama}, and \texttt{hf\_transformers} size against Metal's
advisory working-set hint, which Metal does not enforce;
\texttt{mlx\_lm} lets its KV cache grow on demand. Exact knobs and
API names are listed in Appendix~\ref{sec:framework_settings}.

\paragraph{Memory retained beyond the KV cache.}
Even with a bounded KV cache, MLX's allocator may retain unused buffers
for reuse, keeping the engine's memory footprint high after tensors are
released. Periodic \texttt{clear\_cache} calls or a separate
allocator-cache limit can reduce this retention~\citep{mlxmemorymanagement}.
OS-level controls govern kernel-side GPU memory reclamation and do not
directly manage these allocator caches
(Appendix~\ref{sec:framework_settings}).

\begin{figure}[t]
\centering
\includegraphics[width=\linewidth]{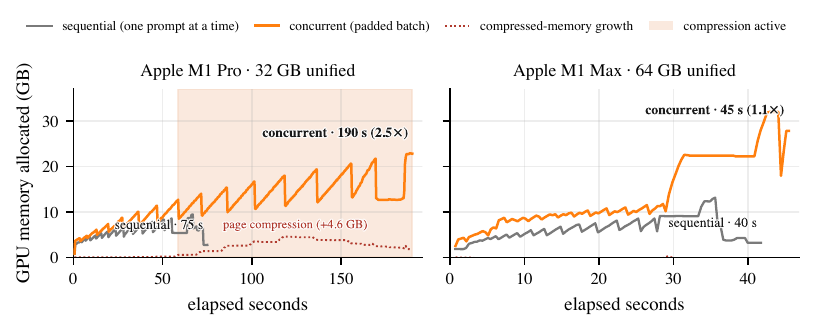}
\caption{\textbf{The workload that fits in 64\,GB thrashes in
32\,GB.} \texttt{mlx\_lm} padded queries and batch-contiguous KV
(30K + 5K + 10 input tokens), sequential vs.\ concurrent, on a 32\,GB M1~Pro
(left) and a 64\,GB M1~Max (right). Concurrent padding triggers
page compression and a $2.5\times$ slowdown on the smaller machine.}
\label{fig:mlx_fragmentation}
\end{figure}

\paragraph{Padded query and KV waste.}
Even when unused buffers are released, padding can inflate the memory
occupied by active requests. Figure~\ref{fig:mlx_fragmentation}
illustrates this effect in \texttt{mlx\_lm}, whose batching layout pads
both query tensors and KV storage. The controlled workload uses three
prompts of 30K, 5K, and 10 input tokens, run first sequentially then
concurrently. In concurrent mode the query tensor
has shape $[3,30000]$ and live K/V uses
$[3,\text{heads},30000,\text{head\_dim}]$: actual tokens total
35{,}010, but both paths span 90{,}000 token positions, wasting 61\%
of query compute and KV memory I/O on padding. The performance impact
also depends on available memory headroom. On the 32\,GB
M1~Pro (left), the wasted memory raises system memory pressure enough
to trigger macOS page compression of active KV cache pages. Every
decode step then pays decompression before attention and
recompression after, and
wall time grows $2.5\times$ relative to sequential execution.
On the 64\,GB M1~Max (right), the same workload fits without triggering
compression, and the concurrent penalty is negligible. The
threshold is not a hard OOM but a deployment cliff: identical code
degrades silently on smaller machines with no error signal.

Section~\ref{sec:results_memory} evaluates system memory use and headroom
under load; Appendix~\ref{sec:appendix_qwen38_27b} extends the audit to
the memory ceiling with Qwen3.8-27B.

%
%
%
%
%
%

\subsection{D3: Multi-node inference}
\label{sec:multi_node}

\emph{Keeping model weights resident in memory beyond a single
box's capacity requires spanning machines on Apple Silicon: every
unit ships as a sealed single-GPU system.}

Connections between Macs have much lower bandwidth than intra-server
NVLink links, making the parallelization and transport choices
central to efficiency. We first distinguish parallelism strategies, then
compare the strategies and transports available in the audited multi-node
stacks.

\paragraph{Parallelism strategies.}
Tensor parallelism (TP) shards weights across devices; pipeline
parallelism (PP) assigns layer ranges per device.

\paragraph{Audit across Apple Silicon stacks.}
EXO~\citep{exo}, an orchestrator outside the single-machine roster,
and \texttt{mlx\_lm}~\citep{mlxdistributed} provide TP over
Thunderbolt~5 RDMA. We compare them with \texttt{llama.cpp}'s
RPC backend~\citep{llamacpprpc}, whose macOS path uses TCP and
executes pipeline stages sequentially (\S\ref{sec:results_d3}).
Backend details are in Appendix~\ref{sec:framework_settings}.
\texttt{vllm-metal} also exposes PP over the
MLX ring backend,\footnote{\url{https://github.com/vllm-project/vllm-metal/pull/427}}
but loads the full model on each node before splitting layers;
we do not evaluate this path.

\section{Workload and Datasets}
\label{sec:dataset}

Guided by these desiderata, \ourbench targets the single-user agent
workload described in \S\ref{sec:intro}: one task fans out to 4--16
parallel requests against a local server. Agent requests carry long
contexts but short replies, so every tool round trip pays TTFT and
end-to-end latency sets the duration of the turn. To compare
single-request and concurrent serving, we sweep concurrency at 1, 8,
and 16 across two prompt splits (Table~\ref{tab:dataset}).

\paragraph{Model selection.}
The three models (Qwen3-0.6B, Qwen3.5-0.8B, Gemma-4-E4B-it) are
small by design. Qwen3-0.6B~\citep{qwen3} is the common model across
all nine stacks on the 64\,GB Apple M5~Pro; the other
two deliberately exercise newer architectures (hybrid linear
attention, multimodality), where per-stack support gaps are
themselves the D1 coverage signal
(\S\ref{sec:attention_variants}). Small models reduce the contribution
of weights to total memory use, making concurrency-dependent runtime
allocations easier to examine. The larger dense/MoE studies
(\S\ref{sec:results_large_models}, Appendix~\ref{sec:appendix_large_models}) and the 35B multi-node
comparison (\S\ref{sec:results_d3}) extend the evaluation to larger models.
The fidelity task remains discriminative at 0.6B.

\begin{table}[H]
\centering
\small
\setlength{\tabcolsep}{4pt}
\captionsetup{skip=3pt}
\setlength{\belowcaptionskip}{0pt}
\begin{tabular*}{\linewidth}{@{\extracolsep{\fill}}lclc@{}}
\toprule[0.4pt]
\textbf{Split} & $\mathbf{n}$ & \textbf{Sources} & \textbf{Token range} \\
\midrule[0.3pt]
Chat  & 100 & OpenOrca, CNN/DailyMail (4 bins) & 21--3.5K \\
Agent & 100 & BFCL V3, Hermes, ClawsBench & 0.3K--8.9K \\
\bottomrule[0.4pt]
\end{tabular*}
\caption{\textbf{Chat and agent splits exercise different serving regimes.}
\ourbench prompt splits; agent tool calls pre-baked. Tokens
counted with Qwen3 tokenizer.}
\label{tab:dataset}
\end{table}

\paragraph{Speed dataset.}
Chat prompts are drawn from OpenOrca~\citep{mukherjee2023orca} and
CNN/DailyMail~\citep{hermann2015teaching}; agent prompts from
BFCL~V3~\citep{yan2024berkeley},
Hermes~\citep{teknium2025hermes}, and
ClawsBench~\citep{li2026clawsbench}.
The chat split fixes output targets at 64 or 256 tokens; the agent
split generates up to the same 256-token cap without a fixed
target. The size $n=100$ per split is calibrated to
the wall-time budget of the maintenance cadence: the harness
enforces a 1\,h per-framework wall-clock cap. We do not score task
accuracy on these speed splits.
The prompt set, sampling seed, and dataset versions are held fixed
across dated snapshots so that performance and fidelity deltas
reflect framework changes rather than dataset drift. Three stacks whose
default context windows sit below the agent split's longest
prompts run with a raised context on that split
(Appendix~\ref{sec:framework_settings}). The main serving splits omit two regimes
that local agents will increasingly hit: very long contexts (above
10K tokens) and long-form generation (above 1K output tokens);
both are scope choices for this snapshot.

\paragraph{Serving protocol.}
Each level follows three warmup requests, then issues requests in a
closed loop with at most $c$ in flight. Payloads set temperature to
zero and request non-thinking generation. Output throughput is
total generated tokens divided by the measured sweep's wall time.
TTFT runs from HTTP submission to the first streamed content or
reasoning output. Latency summaries cover completed requests;
failed requests still contribute to sweep wall time.
Appendix~\ref{sec:framework_settings} details token accounting and
timeouts.

\paragraph{Fidelity dataset.}
Fidelity is measured on a separate dataset: the GMRID supply-chain
incident-classification task~\citep{llmsattheedge2025}
($N{=}1146$, 8 classes, 0-shot and 5-shot).
\S\ref{sec:fidelity} describes the metric and reports per-stack F1.

\section{Human Maintainers and the Maintainer Agent}
\label{sec:fidelity_pillar}
\label{sec:maintainer_agent}

\ourbench targets a verified snapshot every two weeks. A Claude Code
agent updates frameworks, runs the benchmark, and proposes fixes
under released instructions.\footnote{\href{https://github.com/WindChimeRan/SiliconBench/blob/616aa51c450383ee9309d2149d6765f9bd117119/.claude/skills/weekly-bench/SKILL.md}{Maintainer instructions at benchmark commit \texttt{616aa51c}.}}
Its write allowlist covers framework adapters and model profiles;
workloads, scoring, and aggregation remain maintainer-controlled.
Maintainers review agent changes and community PRs before rerunning
official results. Appendix~\ref{sec:framework_settings} records
provenance limits.

\label{sec:maintenance_casestudy}
Two incidents from five early M2~Max benchmark runs (2026-04-11 through
2026-04-23) illustrate the workflow; these runs do not establish a
sustained publication cadence.

\paragraph{Correlated failures from a shared dependency.}
The 2026-04-21 update from MLX 0.31.1 to 0.31.2 broke three
frameworks: \texttt{mlx\_lm} hit a stream error,
\texttt{vllm-metal} failed to compile against removed Metal APIs,
and \texttt{vllm-mlx} returned zero tokens. The agent pinned
\texttt{mlx\_lm} and pulled available upstream fixes. Two stacks
recovered in the same cycle; \texttt{vllm-mlx} remained broken.

\paragraph{A template error corrupts fidelity.}
The agent diagnosed that \texttt{ollama}'s bare-GGUF model import
omitted Qwen3's ChatML delimiters, causing ${\sim}$92\% parse
failures on the classification task. Switching to a registry pull
(where the manifest includes the chat template and stop tokens)
resolved the gap.

\section{Benchmark Results}
\label{sec:baselines}

\subsection{Speed: throughput and latency}
\label{sec:results_speed}

\subsubsection{Nine-engine Apple Silicon audit}
\label{sec:results_d1}

\begin{figure}[H]
\centering
\includegraphics[width=\linewidth]{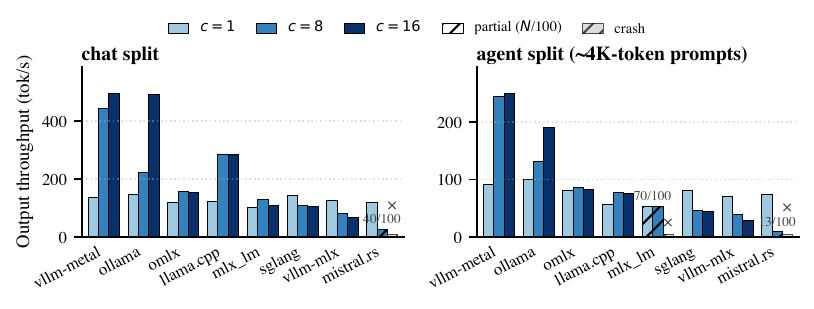}
\caption{\textbf{Similar chat throughput masks different performance
on agent workloads.} Output throughput at $c{=}1$, 8, 16 on the chat
(left) and agent (right) splits, Apple M5~Pro; note the different
y scales. Stacks sorted by agent $c{=}8$. Hatched
colored bars: partial success ($N$/100); gray hatched stubs:
crashed (``$\times$'').
Qwen3-0.6B BF16, $n{=}100$. \texttt{hf\_transformers} excluded:
no completed agent run within the 1\,h cap.}
\label{fig:d1_throughput_scaling}
\end{figure}

We benchmark Qwen3-0.6B BF16 across nine systems on one Apple
M5~Pro (64\,GB, macOS~26.6) at concurrency 1, 8, and 16
(Figure~\ref{fig:d1_throughput_scaling}). \texttt{hf\_transformers}
provides a PyTorch MPS baseline on chat only; the remaining systems
use their native MLX or Metal paths. Appendix~\ref{sec:framework_settings}
records campaign dates and settings. We examine throughput and
TTFT, then request completion and model coverage.

On Qwen3-0.6B, engines with the same audited batching capabilities exhibit
different concurrency scaling. Both \texttt{vllm-metal} and
\texttt{hf\_transformers} support packed queries, paged KV, and mixed steps,
yet their chat throughput scales by $3.65\times$ and $1.23\times$,
respectively, from $c{=}1$ to $c{=}16$. \texttt{vllm-metal} also scales
$2.71\times$ on agent and is the only stack above $2\times$ on
\emph{both} splits.
Every padded or serial-prefill path either gains
less than 30\%, regresses, or fails by $c{=}16$ on chat; on agent,
only \texttt{omlx} stays near its single-stream rate.
At chat $c{=}16$, \texttt{vllm-metal} and \texttt{ollama} finish
within $1\%$, while \texttt{vllm-metal} leads the agent split by
$32\%$. The longer prompts separate the implementations.

First-token latency also separates the stacks
(Figure~\ref{fig:d1_ttft}, Appendix~\ref{sec:appendix_ttft}).
Only \texttt{vllm-metal} combines sub-quarter-second median TTFT
with increasing agent throughput at $c{=}16$ ($219$~ms).
\texttt{sglang} also admits requests quickly ($655$~ms), but its
agent throughput falls $80.6\to45.3$~tok/s. The other agent
survivors exceed one second.

\phantomsection
\label{sec:completion_counts}
Six of the nine stacks complete at least 90/100 requests cleanly
at every level on both splits (Table~\ref{tab:scorecard}).
\texttt{mistral.rs} degrades at $c{=}8$ and crashes at $c{=}16$;
\texttt{mlx\_lm} collapses to 2/100 at agent $c{=}16$; and
\texttt{hf\_transformers} exceeds the campaign's 1\,h wall-clock
cap on agent. These counts require interpretation through
\texttt{finish\_reason}: \texttt{mlx\_lm}'s 70/100 at agent
$c{=}1$ and $c{=}8$ reflects legitimate tool calls parsed to a
structured field with zero content tokens, which the harness
scores as failures. Its $c{=}16$ collapse is real. Completion
alone is insufficient: two of the six surviving stacks fail the
fidelity check (\S\ref{sec:results_fidelity}).

Model coverage supplies the other part of D1. Five stacks serve
both Qwen3.5-0.8B and Gemma-4-E4B-it; \texttt{hf\_transformers}
completes only the Gemma chat split, while \texttt{ollama},
\texttt{mistral.rs}, and Apple \texttt{sglang} serve neither
release. \texttt{vllm-metal} leads throughput at $c{=}16$ on both
splits for both newer models (Appendix~\ref{sec:appendix_speed_extra}).
These smaller rosters assess support for the newer architectures;
the nine-engine scaling comparison above is specific to Qwen3-0.6B.

\subsubsection{Concurrent serving of larger models}
\label{sec:results_large_models}

\begin{figure}[H]
\centering
\includegraphics[width=\linewidth]{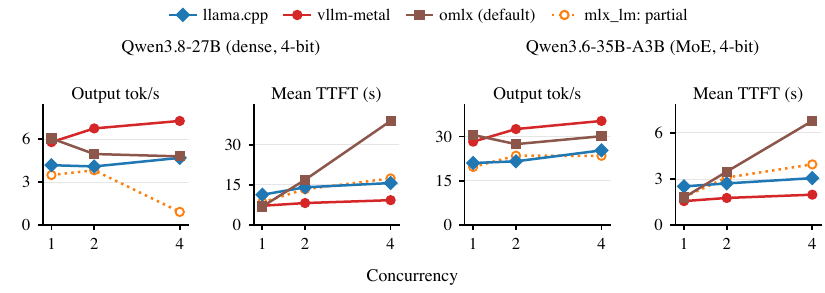}
\caption{\textbf{The MoE throughput gap narrows, while TTFT still separates
the serving paths.} Agent prompts, 100 requests per level; \texttt{omlx}
uses its default SSD-backed prefix cache. Hollow \texttt{mlx\_lm} markers
denote partial runs (23--31/100 on dense, 38--40/100 on MoE);
\texttt{llama.cpp} completes 99/100 and the other engines 100/100.
TTFT averages completed requests, so partial runs cover different
subsets; throughput counts completed output tokens over full sweep
wall time.}
\label{fig:large_model_serving_main}
\end{figure}

To examine concurrent serving beyond the small-model audit, we evaluate
four selected engines on Qwen3.8-27B (dense)
and Qwen3.6-35B-A3B (MoE, 3B active parameters). We use nominal 4-bit
weights on the same 64\,GB M5~Pro and agent prompts at $c{=}1/2/4$
(Figure~\ref{fig:large_model_serving_main}).

The throughput comparison is sensitive to cache configuration. At
$c{=}4$ on the MoE workload, \texttt{vllm-metal}'s measured advantage
ranges from $4\%$ to $17\%$ across the archived \texttt{omlx} cache
configurations.

On the dense 27B model, \texttt{vllm-metal}'s throughput rises with
concurrency, and it has the lowest mean TTFT at $c{=}4$. On MoE,
the throughput gap narrows: \texttt{omlx} leads at $c{=}1$, while
\texttt{vllm-metal} maintains lower TTFT under load.

Each configuration was measured once, using later builds and different
GGUF/MLX quantization formats. Appendix~\ref{sec:appendix_large_model_serving}
provides request latency, configuration details, and cache sensitivity.

\subsubsection{The CUDA reference track: DGX Spark}
\label{sec:results_dgxspark}

\begin{figure}[H]
\centering
\includegraphics[width=\linewidth]{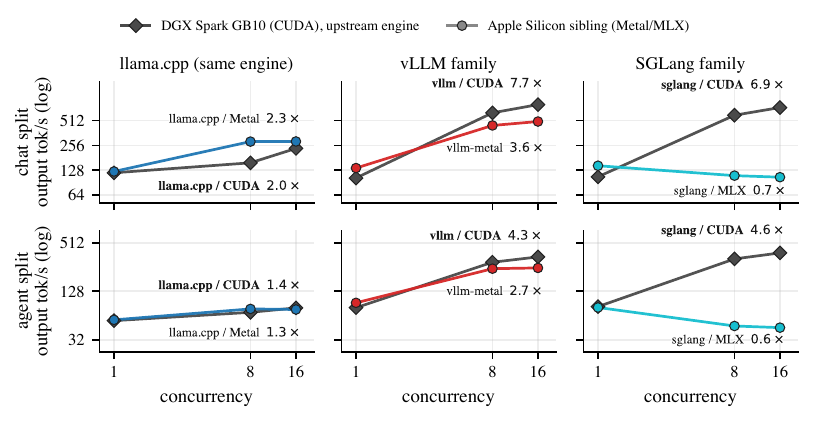}
\caption{\textbf{CUDA vLLM and SGLang sustain stronger concurrency scaling on Qwen3-0.6B.}
Bridge pairs: the three engine families shared between the Apple
M5~Pro roster and the DGX Spark reference track, same workload
(Qwen3-0.6B BF16, $n{=}100$). Columns are engine families, rows
are splits; both axes log. Gray diamonds: upstream engine on
GB10/CUDA; colored circles: the Apple sibling (colors match
Figures~\ref{fig:d2_trajectory}
and~\ref{fig:d3_memory_profile}). Line-end labels give $c{=}1\to16$
scaling.}
\label{fig:bridge_pairs}
\end{figure}

The three shared engine families run on DGX Spark, a CUDA
unified-memory desktop, with identical prompts, weights, and
concurrency grid (Figure~\ref{fig:bridge_pairs}). On Qwen3-0.6B,
CUDA vLLM and SGLang reach $4.3$--$7.7\times$ scaling across the
two splits with flat TTFT. \texttt{llama.cpp}'s agent throughput
flattens on both platforms, while its Metal build finishes ahead
on chat at $c{=}16$. These observations demonstrate serving
headroom, but hardware, builds, and settings differ, including
\texttt{llama.cpp}'s four Apple server slots. The comparison does
not isolate their individual contributions. The track covers
speed and latency only; Appendix~\ref{sec:appendix_dgxspark}
provides full results, including throughput on both newer model releases.

\subsection{Memory discipline under load}
\label{sec:results_memory}

\begin{figure}[H]
\centering
\includegraphics[width=\linewidth]{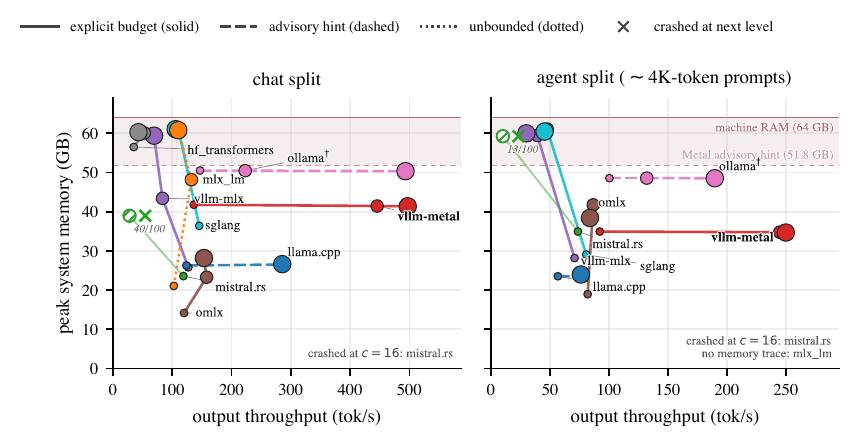}
\caption{\textbf{Completing every request does not ensure memory discipline.}
Throughput and peak system memory for Qwen3-0.6B BF16 on Apple
M5~Pro, with 100 requests per split at concurrency 1, 8, and 16
(increasing marker size). \texttt{sglang} and \texttt{vllm-mlx}
complete all requests despite high memory use; \texttt{vllm-metal}
scales throughput at nearly constant memory. Filled markers
indicate $\geq$90/100 completions; hollow markers indicate partial
runs labeled $N$/100. $^{\dagger}$: fails five-shot fidelity
(Table~\ref{tab:fidelity_qwen06}).}
\label{fig:d2_trajectory}
\end{figure}

Figure~\ref{fig:d2_trajectory} compares throughput with peak system
memory while serving; Appendix~\ref{sec:appendix_memory} supplies
within-run traces. \texttt{vllm-metal} and \texttt{llama.cpp} keep
nearly flat memory use as load increases; the former also scales
throughput, with agent peak system memory of $34.7$--$34.9$\,GB.
\texttt{omlx} completes every request while its memory use grows
with demand but stays below the advisory hint. Flat memory use alone
is insufficient to preserve headroom: \texttt{ollama} holds
94--97\% of the Metal advisory hint across the two splits,
starting at $c{=}1$. The hint marks lost headroom rather than a
failure threshold (\S\ref{sec:memory_management}). This chat
throughput co-leader also fails 5-shot fidelity
(\S\ref{sec:results_fidelity}); a speed-only ranking registers
neither cost.

Declared budgets also fail to keep system memory use low.
\texttt{mistral.rs}, \texttt{vllm-mlx}, and \texttt{sglang} declare
explicit budgets yet approach physical RAM as throughput falls
(Table~\ref{tab:features}, Figure~\ref{fig:d2_trajectory}).
\texttt{mistral.rs} reaches $59.3$\,GB at agent $c{=}8$ while
completing only 13/100 requests. \texttt{sglang} and
\texttt{vllm-mlx} each complete all 600 requests despite high memory
usage, which success counts alone do not reveal. Unbounded
\texttt{mlx\_lm} nearly triples its chat footprint
for a $7\%$ throughput gain. A declared budget is a configuration
knob; \textbf{D2} requires control over the engine's total memory footprint.
The separate 8-bit Qwen3.8-27B study extends this memory audit to
larger weights (Appendix~\ref{sec:appendix_qwen38_27b}).

\needspace{6\baselineskip}
\subsection{Fidelity against the NVIDIA reference}
\label{sec:results_fidelity}
\label{sec:fidelity}

\definecolor{outlierbg}{gray}{0.88}
\begin{table}[H]
\centering
\fontsize{9}{10.5}\selectfont
\setlength{\tabcolsep}{3pt}
\setlength{\abovecaptionskip}{3pt}
\setlength{\belowcaptionskip}{0pt}
\renewcommand{\arraystretch}{1.05}
\begin{tabular*}{\textwidth}{@{\extracolsep{\fill}}lcccccc@{}}
\toprule[0.4pt]
& \multicolumn{2}{c}{\textbf{Qwen3-0.6B}} & \multicolumn{2}{c}{\textbf{Qwen3.5-0.8B}} & \multicolumn{2}{c}{\textbf{Gemma-4-E4B-it}} \\
\cmidrule[0.4pt](lr){2-3} \cmidrule[0.4pt](lr){4-5} \cmidrule[0.4pt](lr){6-7}
\textbf{Stack} & 0-shot & 5-shot & 0-shot & 5-shot & 0-shot & 5-shot \\
\midrule[0.4pt]
\texttt{vllm-nv} (ref) & .4094 & .7364 & .6953 & .5949 & .8511 & .9128 \\
\texttt{ollama} & .4173 & \cellcolor{outlierbg}.4462 & --- & --- & --- & --- \\
\texttt{llama.cpp} & .4033 & .7377 & .7003 & .5940 & .8497 & .9107 \\
\texttt{hf\_trans.} & .3995 & .7299 & --- & --- & --- & --- \\
\texttt{mlx\_lm} & .3970 & .7346 & .6942 & .5903 & .8493 & .9108 \\
\texttt{vllm-metal} & .3956 & .7423 & .6947 & .6081 & .8512 & .9106 \\
\texttt{sglang} & .3955 & .7337 & --- & --- & --- & --- \\
\texttt{omlx} & .3944 & .7379 & .6981 & .5929 & .8539 & .9120 \\
\texttt{vllm-mlx} & \cellcolor{outlierbg}.3592 & \cellcolor{outlierbg}.7183 & .7006 & .5922 & .8530 & .9107 \\
\bottomrule[0.4pt]
\end{tabular*}
\caption{\textbf{Six of eight stacks match on Qwen3-0.6B; all five
match on each newer model.} Fidelity (BF16), weighted F1 on
GMRID~\citep{llmsattheedge2025}: generations on the Apple M5~Pro
(2026-08-21 builds), scored against the \texttt{vllm-nv} reference
on an NVIDIA A100. \colorbox{outlierbg}{Shaded}: outside the
$\sim$1.5pp band around the reference for that model and shot count.
---: not measured; newer-model coverage is limited to five Apple stacks.}
\label{tab:fidelity_qwen06}
\end{table}

We assess task-level fidelity using weighted F1 on GMRID against
a vLLM reference on an NVIDIA A100, with identical weights and
precision; token-level comparison was dominated by chat-template
differences. Table~\ref{tab:fidelity_qwen06} covers eight
Qwen3-0.6B stacks (\texttt{mistral.rs} was too slow to finish)
and five on each newer release (setup in
Appendix~\ref{sec:appendix_crossmodel}). Six Qwen3-0.6B stacks
lie within roughly 1.5pp of the reference on both 0-shot and
5-shot; all five on each newer model lie within 1.4pp. This
observed agreement band is not a formal noise estimate, so we
do not rank stacks within it.

On Qwen3-0.6B, \texttt{ollama} lies within the 0-shot reference
band but gains little with 5 shots, falling ${\sim}$30pp
behind stacks benefiting from in-context examples
(Table~\ref{tab:fidelity_qwen06}). Its chat template's handling
of multiple examples is the likely cause, though we did not
isolate the root cause. This differs from the April template-import
bug (\S\ref{sec:maintenance_casestudy}). On tool-call prompts,
it also streams ${\sim}254$ tokens with empty content (verified
in output sidecars), so its agent throughput partly counts
content-free tokens.

The Qwen3-0.6B result may indicate an implementation bug:
\texttt{vllm-mlx} scores 5.0 percentage points below the zero-shot
reference despite zero request or parse errors. Its scores on both
newer models fall within the reference band.

\subsection{Multi-node decode throughput}
\label{sec:results_d3}

We measure single-request throughput on Qwen3.5-35B-A3B at 8-bit
on one or two M4~Pro Mac mini (64\,GB each) connected by
Thunderbolt~5 with RDMA. The model fits on a single node, so the
comparison isolates multi-node engine efficiency from the
capacity-unlock case. Table~\ref{tab:multi_node} compares
\texttt{llama.cpp}'s PP-over-TCP path with \texttt{mlx\_lm} and
EXO using TP over RDMA (\S\ref{sec:multi_node}).

\begin{table}[H]
\centering
\small
\setlength{\tabcolsep}{5pt}
\rowcolors{2}{tablerow}{white}
\begin{tabular}{@{}>{\columncolor{white}[0pt][\tabcolsep]}lcccccc>{\columncolor{white}[\tabcolsep][0pt]}c@{}}
\toprule
\rowcolor{white}
                          & \multicolumn{3}{c}{\textbf{Decode TPS at ctx}} & \multicolumn{3}{c}{\textbf{Prefill (s) at ctx}} & \textbf{2-node} \\
\cmidrule(lr){2-4} \cmidrule(lr){5-7}
\rowcolor{white}
\textbf{Stack}            & \textbf{512} & \textbf{4096} & \textbf{16384} & \textbf{512} & \textbf{4096} & \textbf{16384} & \textbf{speedup} \\
\midrule
\rowcolor{white}
\multicolumn{8}{@{}l}{\emph{Single node}} \\
\texttt{llama.cpp}        & 40.0  & 39.0  & 35.7  & 0.88 & 5.63  & 25.01 & ---    \\
\texttt{mlx\_lm}          & 49.3  & 48.0  & 43.7  & 1.32 & 4.96  & 22.84 & ---    \\
EXO                       & 66.8  & 63.9  & 57.0  & 1.24 & 4.47  & 18.83 & ---    \\
\midrule
\rowcolor{white}
\multicolumn{8}{@{}l}{\emph{Two nodes}} \\
\texttt{llama.cpp} (PP)   & 32.4  & 31.2  & 29.9  & ---  & ---   & ---   & 0.80--0.84$\times$ \\
\texttt{mlx\_lm} (TP)     & 63.3  & 62.7  & 57.7  & 0.65 & 2.89  & 13.63 & 1.28--1.32$\times$ \\
EXO (TP)                  & 87.0  & 85.1  & 78.3  & 0.57 & 2.48  & 11.22 & 1.30--1.37$\times$ \\
\bottomrule
\end{tabular}
\caption{\textbf{TP over RDMA scales; PP over TCP regresses.}
Throughput on Qwen3.5-35B-A3B 8-bit, 512 output tokens. 1--2
M4~Pro Mac mini (64\,GB), TB5+RDMA. \texttt{llama.cpp}: PP over
TCP (no RDMA on macOS).}
\label{tab:multi_node}
\end{table}

TP over RDMA scales modestly on decode
(Table~\ref{tab:multi_node}, rightmost column) and better on
prefill, which is compute-bound: roughly $2\times$ at short
contexts and $1.7$--$1.8\times$ at 4K tokens for both stacks.
EXO's absolute lead also includes its $35\%$ single-node advantage
over \texttt{mlx\_lm}. \texttt{llama.cpp}'s sequential TCP pipeline
regresses $16$--$20\%$ on two nodes.

\needspace{12\baselineskip}
\section{Discussion}
\label{sec:discussion}

\paragraph{Agentic desktop inference requires varlen serving.}
Packed varlen queries, paged KV, and mixed steps should be the
baseline design for agentic desktop serving. Their implementation
must also be judged by measured throughput and latency
(\S\ref{sec:results_d1}).

\paragraph{The audit gates.}
The audit gates establish minimum usability under the tested workloads.
The remaining differences concern how engines provide that usability:
their concurrency scaling, memory policies, and ability to distribute
inference.
\texttt{llama.cpp}, \texttt{vllm-metal}, and \texttt{omlx} pass
all three gates in this snapshot (Table~\ref{tab:scorecard}). The
fidelity exclusions remain
model- or shot-specific (\S\ref{sec:results_fidelity}).

\begin{table}[H]
\centering
\small
\setlength{\tabcolsep}{5pt}
\rowcolors{2}{tablerow}{white}
\begin{tabular}{@{}>{\columncolor{white}[0pt][\tabcolsep]}lllcc>{\columncolor{white}[\tabcolsep][0pt]}c@{}}
\toprule
\rowcolor{white}
\textbf{Stack} & \textbf{Speed} & \textbf{Memory} & \textbf{Fidelity} & \textbf{Models} & \textbf{All gates} \\
\midrule
\texttt{llama.cpp}   & fast chat, mid agent    & low, flat   & ref.\ band & 3/3 & \checkmark \\
\texttt{vllm-metal}  & scales to $c{=}16$      & flat budget & ref.\ band & 3/3 & \checkmark \\
\texttt{omlx}        & mid; flat under load    & grows, bounded & ref.\ band & 3/3 & \checkmark \\
\midrule
\texttt{sglang}      & declines under load     & sawtooth to 61\,GB & ref.\ band & 1/3 & \\
\texttt{ollama}      & fast, esp.\ chat        & high plateau & fails 5-shot & 1/3 & \\
\texttt{mlx\_lm}     & collapses agent $c{=}16$ & climbs (chat) & ref.\ band & 3/3 & \\
\texttt{mistral.rs}  & partial $c{=}8$, crash $c{=}16$ & near-RAM & excluded & 1/3 & \\
\texttt{vllm-mlx}    & declines, both splits   & climbs, both splits & outlier & 3/3 & \\
\texttt{hf\_trans.}  & chat-only (1\,h cap)     & 56--60\,GB & ref.\ band & 2/3 & \\
\bottomrule
\end{tabular}
\caption{\textbf{Three stacks pass all three audit gates.}
Speed, memory, and fidelity summarize Qwen3-0.6B. The three minimum
audit gates are completion
($\geq$90/100 successful requests in the harness) at every concurrency level on both
splits, reference-band fidelity, and serving all three model
releases (\textbf{Models}). Speed is reported, not gated: a slow
stack is still usable, an absent one is not.
\texttt{mlx\_lm}'s agent run carries no memory trace.}
\label{tab:scorecard}
\end{table}

\paragraph{Memory and multi-node remain open.}
Architecture readiness does not close the other two desiderata.
The D2 measurements distinguish flat footprints and bounded demand-driven
growth from paths that approach physical capacity or crash. Adaptation to
changing foreground memory demand was not evaluated.
EXO and \texttt{mlx\_lm} increase decode throughput on two nodes using
tensor parallelism over RDMA; \texttt{llama.cpp}'s sequential pipeline
reduces it.

\section{Limitations}

The main serving audit uses one Apple M5~Pro; the DGX Spark
reference covers three engine families on throughput and latency
only. Small models anchor the main audit, while larger dense/MoE
and multi-node studies cover selected engines and configurations.
The larger-model follow-up uses later builds and different
quantization formats. Fidelity rests on a single classification
task, and we report no formal run-to-run variance estimates.
The multi-node study uses two machines and a model that also
fits on one.
Appendix~\ref{sec:appendix_limitations} gives details.

\section{Conclusion}

\ourbench shows why local serving requires joint evaluation of speed,
memory, and fidelity. Declared features do not guarantee concurrency
scaling, and successful requests do not guarantee memory headroom or
reference fidelity. Multi-node scaling likewise varies across the tested
combinations of parallelism and transport. Fixed workloads and reviewed
maintenance provide a basis for reassessing these findings as serving
software evolves.


\bibliographystyle{plainnat}
\bibliography{ref}

\clearpage
\appendix
\etocdepthtag{appendix}

\begingroup
\small
\setlength{\parskip}{2pt}
\etocsettagdepth{main}{none}
\etocsettagdepth{appendix}{subsection}
\etocsettocstyle{\section*{Appendix contents}}{}
\etocsetstyle{section}{}{}
  {\noindent\makebox[2em][l]{\etocnumber}\etocname
   \nobreak\dotfill\makebox[1.5em][r]{\etocpage}\par}{}
\etocsetstyle{subsection}{}{}
  {\noindent\hspace*{1em}\makebox[2.5em][l]{\etocnumber}\etocname
   \nobreak\dotfill\makebox[1.5em][r]{\etocpage}\par}{}
\tableofcontents
\endgroup

\section{Experimental Setup and Framework Settings}
\label{sec:framework_settings}

\begin{table}[H]
\centering
\small
\setlength{\tabcolsep}{4pt}
\renewcommand{\arraystretch}{1.15}
\begin{tabularx}{\linewidth}{@{}>{\raggedright\arraybackslash}X>{\raggedright\arraybackslash}p{1.1in}rr>{\raggedright\arraybackslash}p{0.85in}@{}}
\toprule
\textbf{Role} & \textbf{Machine} & \textbf{RAM} & \textbf{Hint} & \textbf{OS / stack} \\
\midrule
Main track & Apple M5~Pro & 64\,GB & 51.84\,GB & macOS~26.6 \\
Left-pad study & Apple M1~Pro & 32\,GB & 24.96\,GB & macOS \\
Left-pad study & Apple M1~Max & 64\,GB & 48\,GB & macOS \\
Multi-node & $2\times$ M4~Pro Mac mini & 64\,GB each & --- & macOS~26 \\
CUDA reference & DGX Spark (GB10) & 128\,GB & --- & Linux / CUDA \\
Maintenance studies & Apple M2~Max & 64\,GB & 48\,GB & macOS~26 \\
Fidelity reference (\S\ref{sec:results_fidelity}) & $1\times$ NVIDIA A100 & --- & --- & vLLM \\
\bottomrule
\end{tabularx}
\caption{\textbf{All main-track stacks share one M5~Pro.}
All main-track speed, memory, and fidelity-generation numbers come
from this machine, including the extra-model tables and larger-model follow-ups. RAM is
unified on the Apple and Spark machines. The multi-node pair uses
Thunderbolt~5 with RDMA. Hint is Metal's
\texttt{recommendedMaxWorkingSetSize}.}
\label{tab:hardware}
\end{table}

\paragraph{Main-track provenance.}
The Qwen3-0.6B campaign ran on 2026-08-22, with repeats on
August~25--26 against the same installed builds. The August~25
repeats followed a fix to means of per-request rates; aggregate
output throughput was unaffected. August~26 added telemetry for
\texttt{sglang} and \texttt{omlx} on both splits and
\texttt{vllm-mlx} on agent. \texttt{sglang}'s declared
\texttt{transformers} pin was restored before the repeat.
\texttt{mlx\_lm}'s agent cells retain the August~25 measurement:
the telemetry repeat exceeded the wall-clock cap after its Metal
OOM collapse, leaving memory unmeasured. \texttt{omlx} starts each
level with a fresh server and SSD prefix-cache directory; its
in-memory tier is disabled (\texttt{--hot-cache-max-size 0}).
The retained results, prompts, and adapters are pinned to benchmark
commit \texttt{616aa51c}, with file hashes in the paper's
\texttt{reproducibility/paper\_snapshot.json}.\footnote{\url{https://github.com/WindChimeRan/SiliconBench/tree/616aa51c450383ee9309d2149d6765f9bd117119}}
The historical files lack a complete engine-version manifest and
exact agent model IDs; missing fields are not filled from current
installations.

\paragraph{Timing and token accounting.}
Warmup time is excluded. Requests have a 300\,s HTTP timeout,
and the campaign driver imposes a 1\,h per-framework wall-clock
cap. Output counts use server-reported token usage when available;
otherwise the client falls back to streamed delta counts and flags
the missing usage. Failed requests contribute to sweep wall time
but not to latency percentiles or the output-token numerator.
The reported ITL p50 is the median of request-level mean decode
intervals, not a median over individual token gaps; streamed token
bundling limits its comparability. The 27B table states this caveat
for its \texttt{omlx} cells.

\paragraph{Per-stack context windows on the agent split.}
Agent prompts reach $8{,}866$ input tokens at the longest, averaging ${\sim}4.3$K
(Table~\ref{tab:dataset}). Three stacks default to context windows
below this (\texttt{llama.cpp}, \texttt{ollama},
\texttt{vllm-metal}), so the harness raises their per-slot context
to 16384 on the agent split; without this they reject
(\texttt{vllm-metal}) or silently truncate (\texttt{llama.cpp},
\texttt{ollama}) the longest prompts. The remaining stacks serve
Qwen3-0.6B's native 40K context unchanged. Both Apple splits use four
\texttt{llama.cpp} server slots (\texttt{--parallel 4}); higher client
concurrency queues requests beyond those slots.

\paragraph{Roster.}
A tenth stack, \texttt{inferrs}, was benchmarked in early weekly
runs and retired on 2026-07-03 after 69 days without upstream
commits or releases; it appears in no reported result.

\paragraph{Multi-node backends.}
EXO and \texttt{mlx\_lm} use JACCL for TP over Thunderbolt~5 RDMA;
EXO's MLX fork includes RDMA-specific fixes. Qwen3.5 supports TP in
\texttt{mlx\_lm}; \texttt{llama.cpp} uses RPC over TCP, without
async or RDMA support on macOS.

\paragraph{Configuration flags.}
Per-stack serve scripts and environment variables are documented in
the released harness repository. Key configuration flags used by
the harness:
\texttt{VLLM\_METAL\_MEMORY\_FRACTION} (vllm-metal KV budget,
\S\ref{sec:memory_management}),
\texttt{OLLAMA\_NUM\_PARALLEL} and \texttt{OLLAMA\_CONTEXT\_LENGTH}
(ollama slot partitioning), and the Metal advisory
\texttt{recommendedMaxWorkingSetSize} used by llama.cpp, ollama, and
\texttt{hf\_transformers} for hint-based allocation. OS-level knobs
(\texttt{iogpu.wired\_limit\_mb},
\texttt{iogpu.disable\_wired\_collector},
\texttt{iogpu.dynamic\_lwm}) govern kernel-side GPU memory
reclamation but do not reach into userspace allocator caches.

\needspace{32\baselineskip}
\section{TTFT heatmap}
\label{sec:appendix_ttft}

\begin{figure}[h]
\centering
\includegraphics[width=0.92\linewidth]{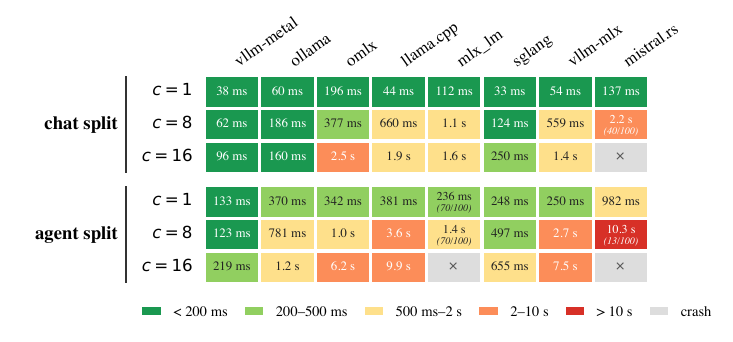}
\caption{\textbf{Only \texttt{vllm-metal} pairs sub-quarter-second
TTFT under load with throughput that scales.} TTFT p50 heatmap;
\texttt{sglang}, the only other stack below one second at
$c{=}16$ on both splits, admits requests fast and then decodes
slowly. Rows are (split, concurrency) pairs; columns are the 8
stacks in the same order as
Figure~\ref{fig:d1_throughput_scaling}
(\texttt{hf\_transformers} excluded: no completed agent run within
the 1~h cap). Italicized counts show
partial-success runs ($N/100$). Qwen3-0.6B BF16 on the Apple
M5~Pro.}
\label{fig:d1_ttft}
\end{figure}
\FloatBarrier

\needspace{20\baselineskip}
\section{Memory traces}
\label{sec:appendix_memory}

\begin{figure}[h]
\centering
\includegraphics[width=\linewidth]{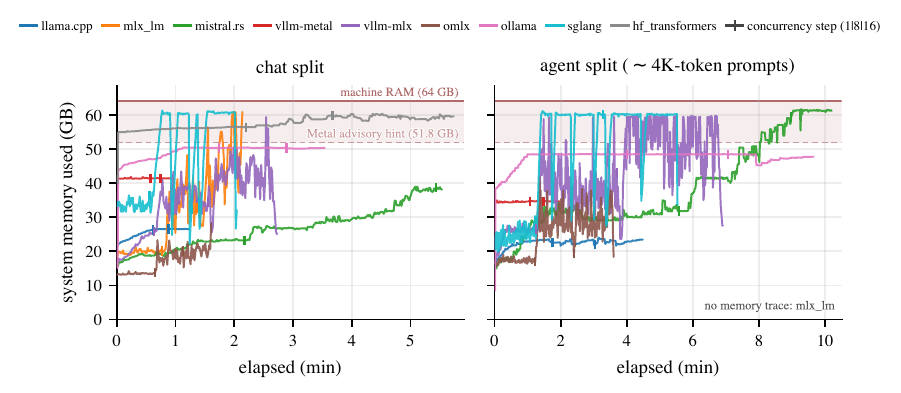}
\caption{\textbf{Memory traces distinguish flat reservations from
growth and allocator sawtooths.} System memory used over elapsed
wall-clock time, sampled once per second with \texttt{metalstat}.
\texttt{omlx}'s independent per-level runs are concatenated;
the other traces span a shared server session. Vertical ticks mark
the boundaries between concurrency levels
$c{=}1{\to}8{\to}16$. The solid rule is machine RAM (64~GB), the
dashed rule the Metal advisory hint (51.8~GB), and the shaded band
between them the low-headroom region
(\S\ref{sec:results_memory}); panel footers list stacks with no
memory trace in this campaign. Qwen3-0.6B BF16 on
the Apple M5~Pro, 100 prompts per level.}
\label{fig:d3_memory_profile}
\end{figure}
\FloatBarrier

\needspace{8\baselineskip}
\section{Cross-model fidelity setup}
\label{sec:appendix_crossmodel}

Table~\ref{tab:fidelity_qwen06} in \S\ref{sec:results_fidelity} reports
all three model releases. The Qwen3.5-0.8B and Gemma-4-E4B-it
evaluations cover five stacks, under the same BF16 protocol as
Qwen3-0.6B: generations on the Apple M5~Pro (2026-08-21 builds),
scored against the \texttt{vllm-nvidia} reference on an NVIDIA
A100. Each score is compared with the reference for the same model
and shot count; unmeasured combinations are marked with dashes.

\FloatBarrier

\section{Per-concurrency serving results for additional model releases}
\label{sec:appendix_speed_extra}

This appendix reports per-stack speed on Qwen3.5-0.8B and Gemma-4-E4B-it,
at concurrency 1, 8, and 16 on the same Apple M5~Pro as the main
track. \texttt{ollama}, \texttt{mistral.rs}, and \texttt{sglang}
could not serve either release; \texttt{hf\_transformers} completed
only the Gemma-4-E4B-it chat split. Agent-table success counts carry
the tool-call serialization caveat of
\S\ref{sec:completion_counts}.

\begin{table}[H]
\centering
\small
\begin{tabular}{@{}lrrrr@{}}
\toprule
\textbf{Stack} & \textbf{Out tok/s} & \textbf{TTFT p50 (ms)} & \textbf{ITL p50 (ms)} & \textbf{Success} \\
\midrule
\rowcolor{gray!16}[0pt][\tabcolsep]\multicolumn{5}{@{}l}{\textit{Concurrency 1}} \\
\texttt{omlx} & 113.4 & 249 & 5.2 & 100/100 \\
\texttt{llama.cpp} & 105.5 & 68 & 8.3 & 100/100 \\
\texttt{vllm-metal} & 100.9 & 64 & 8.8 & 100/100 \\
\texttt{vllm-mlx} & 95.3 & 65 & 9.0 & 100/100 \\
\texttt{mlx\_lm} & 93.9 & 182 & 8.0 & 100/100 \\
\midrule
\rowcolor{gray!16}[0pt][\tabcolsep]\multicolumn{5}{@{}l}{\textit{Concurrency 8}} \\
\texttt{vllm-metal} & 338.7 & 95 & 21.1 & 100/100 \\
\texttt{llama.cpp} & 304.0 & 966 & 12.7 & 100/100 \\
\texttt{vllm-mlx} & 285.1 & 230 & 22.7 & 100/100 \\
\texttt{omlx} & 208.0 & 427 & 32.6 & 100/100 \\
\texttt{mlx\_lm} & 139.1 & 1051 & 32.8 & 100/100 \\
\midrule
\rowcolor{gray!16}[0pt][\tabcolsep]\multicolumn{5}{@{}l}{\textit{Concurrency 16}} \\
\texttt{vllm-metal} & 456.1 & 131 & 30.4 & 100/100 \\
\texttt{vllm-mlx} & 318.0 & 430 & 42.2 & 100/100 \\
\texttt{llama.cpp} & 296.0 & 3064 & 12.9 & 100/100 \\
\texttt{omlx} & 212.6 & 3332 & 34.2 & 100/100 \\
\texttt{mlx\_lm} & 144.4 & 1238 & 83.6 & 100/100 \\
\bottomrule
\end{tabular}
\caption{\textbf{\texttt{vllm-metal} leads Qwen3.5-0.8B chat at $c{=}16$: 456 tok/s, 4.5$\times$ its single-stream rate.} Per-stack speed on the Qwen3.5-0.8B chat split, Apple M5~Pro, 100 prompts per level, sorted by output throughput within each concurrency level. Stacks not evaluated on this model release are omitted.}
\label{tab:speed_qwen35_chat}
\end{table}

\begin{table}[H]
\centering
\small
\begin{tabular}{@{}lrrrr@{}}
\toprule
\textbf{Stack} & \textbf{Out tok/s} & \textbf{TTFT p50 (ms)} & \textbf{ITL p50 (ms)} & \textbf{Success} \\
\midrule
\rowcolor{gray!16}[0pt][\tabcolsep]\multicolumn{5}{@{}l}{\textit{Concurrency 1}} \\
\texttt{omlx} & 77.0 & 441 & 3.4 & 100/100 \\
\texttt{vllm-metal} & 76.2 & 200 & 8.9 & 100/100 \\
\texttt{vllm-mlx} & 67.6 & 342 & 9.9 & 100/100 \\
\texttt{llama.cpp} & 66.9 & 306 & 8.4 & 100/100 \\
\texttt{mlx\_lm} & 45.4 & 504 & 7.5 & 25/100 \\
\midrule
\rowcolor{gray!16}[0pt][\tabcolsep]\multicolumn{5}{@{}l}{\textit{Concurrency 8}} \\
\texttt{vllm-metal} & 257.5 & 119 & 27.9 & 100/100 \\
\texttt{llama.cpp} & 126.9 & 2243 & 17.6 & 100/100 \\
\texttt{vllm-mlx} & 126.0 & 685 & 49.4 & 100/100 \\
\texttt{omlx} & 98.5 & 1248 & 53.1 & 100/100 \\
\texttt{mlx\_lm} & 49.5 & 2599 & 49.4 & 26/100 \\
\midrule
\rowcolor{gray!16}[0pt][\tabcolsep]\multicolumn{5}{@{}l}{\textit{Concurrency 16}} \\
\texttt{vllm-metal} & 297.0 & 181 & 45.6 & 100/100 \\
\texttt{llama.cpp} & 135.3 & 5931 & 16.5 & 100/100 \\
\texttt{vllm-mlx} & 109.4 & 1188 & 130.5 & 100/100 \\
\texttt{omlx} & 92.7 & 5879 & 57.2 & 100/100 \\
\texttt{mlx\_lm} & 41.7 & 3292 & 171.9 & 26/100 \\
\bottomrule
\end{tabular}
\caption{\textbf{\texttt{vllm-metal} leads Qwen3.5-0.8B agent at $c{=}16$: 297 tok/s, 3.9$\times$ its single-stream rate.} Per-stack speed on the Qwen3.5-0.8B agent split, Apple M5~Pro, 100 prompts per level, sorted by output throughput within each concurrency level. Stacks not evaluated on this model release are omitted. Failed counts on the agent split partly reflect tool-call serialization; see the completion-count note (\S\ref{sec:completion_counts}).}
\label{tab:speed_qwen35_agent}
\end{table}

\begin{table}[H]
\centering
\small
\begin{tabular}{@{}lrrrr@{}}
\toprule
\textbf{Stack} & \textbf{Out tok/s} & \textbf{TTFT p50 (ms)} & \textbf{ITL p50 (ms)} & \textbf{Success} \\
\midrule
\rowcolor{gray!16}[0pt][\tabcolsep]\multicolumn{5}{@{}l}{\textit{Concurrency 1}} \\
\texttt{vllm-metal} & 24.3 & 179 & 38.1 & 100/100 \\
\texttt{vllm-mlx} & 24.1 & 239 & 37.2 & 100/100 \\
\texttt{omlx} & 24.1 & 385 & 35.5 & 100/100 \\
\texttt{mlx\_lm} & 22.6 & 303 & 38.5 & 100/100 \\
\texttt{llama.cpp} & 22.4 & 282 & 38.1 & 100/100 \\
\texttt{hf\_transformers} & 14.1 & 329 & 57.7 & 100/100 \\
\midrule
\rowcolor{gray!16}[0pt][\tabcolsep]\multicolumn{5}{@{}l}{\textit{Concurrency 8}} \\
\texttt{vllm-mlx} & 95.0 & 738 & 64.5 & 100/100 \\
\texttt{vllm-metal} & 93.4 & 244 & 76.4 & 100/100 \\
\texttt{mlx\_lm} & 77.1 & 1154 & 79.3 & 100/100 \\
\texttt{llama.cpp} & 77.0 & 3377 & 45.5 & 100/100 \\
\texttt{omlx} & 66.1 & 920 & 103.0 & 100/100 \\
\texttt{hf\_transformers} & 14.7 & 36449 & 57.0 & 100/100 \\
\midrule
\rowcolor{gray!16}[0pt][\tabcolsep]\multicolumn{5}{@{}l}{\textit{Concurrency 16}} \\
\texttt{vllm-metal} & 151.8 & 265 & 84.6 & 100/100 \\
\texttt{vllm-mlx} & 110.6 & 1372 & 124.8 & 100/100 \\
\texttt{mlx\_lm} & 88.5 & 2272 & 145.9 & 100/100 \\
\texttt{llama.cpp} & 81.5 & 9718 & 45.1 & 100/100 \\
\texttt{omlx} & 68.3 & 8884 & 103.4 & 100/100 \\
\texttt{hf\_transformers} & 14.7 & 73550 & 56.4 & 100/100 \\
\bottomrule
\end{tabular}
\caption{\textbf{\texttt{vllm-metal} leads Gemma-4-E4B-it chat at $c{=}16$: 152 tok/s, 6.3$\times$ its single-stream rate.} Per-stack speed on the Gemma-4-E4B-it chat split, Apple M5~Pro, 100 prompts per level, sorted by output throughput within each concurrency level. Stacks not evaluated on this model release are omitted.}
\label{tab:speed_gemma4_chat}
\end{table}

\begin{table}[H]
\centering
\small
\begin{tabular}{@{}lrrrr@{}}
\toprule
\textbf{Stack} & \textbf{Out tok/s} & \textbf{TTFT p50 (ms)} & \textbf{ITL p50 (ms)} & \textbf{Success} \\
\midrule
\rowcolor{gray!16}[0pt][\tabcolsep]\multicolumn{5}{@{}l}{\textit{Concurrency 1}} \\
\texttt{omlx} & 21.6 & 717 & 36.1 & 94/100 \\
\texttt{vllm-metal} & 21.1 & 551 & 41.4 & 93/100 \\
\texttt{vllm-mlx} & 19.4 & 1134 & 37.7 & 94/100 \\
\texttt{llama.cpp} & 18.3 & 1208 & 37.9 & 100/100 \\
\texttt{mlx\_lm} & 17.5 & 826 & 38.7 & 56/100 \\
\midrule
\rowcolor{gray!16}[0pt][\tabcolsep]\multicolumn{5}{@{}l}{\textit{Concurrency 8}} \\
\texttt{vllm-metal} & 52.3 & 905 & 135.9 & 93/100 \\
\texttt{omlx} & 48.3 & 2024 & 124.3 & 94/100 \\
\texttt{vllm-mlx} & 46.7 & 2232 & 124.5 & 94/100 \\
\texttt{llama.cpp} & 44.2 & 9382 & 59.1 & 100/100 \\
\texttt{mlx\_lm} & 23.4 & 2593 & 275.2 & 56/100 \\
\midrule
\rowcolor{gray!16}[0pt][\tabcolsep]\multicolumn{5}{@{}l}{\textit{Concurrency 16}} \\
\texttt{vllm-metal} & 63.6 & 1491 & 227.2 & 93/100 \\
\texttt{omlx} & 50.4 & 15051 & 135.4 & 94/100 \\
\texttt{vllm-mlx} & 49.4 & 2740 & 243.3 & 94/100 \\
\texttt{llama.cpp} & 40.2 & 30180 & 67.8 & 100/100 \\
\texttt{mlx\_lm} & 21.8 & 4402 & 511.4 & 56/100 \\
\bottomrule
\end{tabular}
\caption{\textbf{\texttt{vllm-metal} leads Gemma-4-E4B-it agent at $c{=}16$: 64 tok/s, 3.0$\times$ its single-stream rate.} Per-stack speed on the Gemma-4-E4B-it agent split, Apple M5~Pro, 100 prompts per level, sorted by output throughput within each concurrency level. Stacks not evaluated on this model release are omitted. Failed counts on the agent split partly reflect tool-call serialization; see the completion-count note (\S\ref{sec:completion_counts}).}
\label{tab:speed_gemma4_agent}
\end{table}

\FloatBarrier

\section{Larger-model comparisons}
\label{sec:appendix_large_models}

\subsection{Qwen3.8-27B at the memory ceiling}
\label{sec:appendix_qwen38_27b}

The nine-engine audit uses Qwen3-0.6B; this subsection reports the same
harness at the opposite end of the memory range, serving
Qwen3.8-27B on the same 64\,GB Apple M5~Pro on the agent split at
$c{=}1/2/4$ (Table~\ref{tab:qwen38_27b}). The study evaluates three
of the nine stacks: \texttt{llama.cpp}, \texttt{vllm-metal}, and
\texttt{omlx}.

\begin{table}[h]
\centering
\small
\setlength{\tabcolsep}{4.5pt}
\begin{tabular}{@{}lrrrrr@{}}
\toprule
\textbf{Stack} & \textbf{Out tok/s} & \textbf{TTFT p50 (s)} & \textbf{ITL p50 (ms)} & \textbf{Peak mem (GB)} & \textbf{Success} \\
\midrule
\rowcolor{gray!16}[0pt][\tabcolsep]\multicolumn{6}{@{}l}{\textit{Concurrency 1}} \\
\texttt{llama.cpp} & 3.8 & 8.1 & 104.0 & 51.2 & 100/100 \\
\texttt{vllm-metal} & 3.9 & 7.9 & 108.3 & 53.9 & 100/100 \\
\texttt{omlx} (default) & 4.7 & 6.6 & 102.3 & --- & 100/100 \\
\midrule
\rowcolor{gray!16}[0pt][\tabcolsep]\multicolumn{6}{@{}l}{\textit{Concurrency 2}} \\
\texttt{llama.cpp} & 4.2 & 9.7 & 206.8 & 53.0 & 100/100 \\
\texttt{vllm-metal} & 4.7 & 9.4 & 146.8 & 53.9 & 100/100 \\
\texttt{omlx} (default) & 5.2 & 11.8 & 197.8 & --- & 100/100 \\
\midrule
\rowcolor{gray!16}[0pt][\tabcolsep]\multicolumn{6}{@{}l}{\textit{Concurrency 4}} \\
\texttt{llama.cpp} & 4.7 & 11.1 & 463.0 & 51.1 & 100/100 \\
\texttt{vllm-metal} & 5.2 & 11.7 & 445.1 & 54.0 & 100/100 \\
\texttt{omlx} (default) & 5.9 & 24.1 & 312.2 & --- & 100/100 \\
\bottomrule
\end{tabular}
\caption{\textbf{At the 64\,GB memory ceiling all three evaluated engines complete every request; latency and memory discipline separate them.} Qwen3.8-27B agent split on the 64\,GB Apple M5~Pro, 100 prompts per level at $c{=}1/2/4$. BF16 weights (55.6\,GB) exceed the Metal advisory working-set hint; the comparison uses 8-bit weights: GGUF Q8\_0 for \texttt{llama.cpp}, the mlx-community 8-bit conversion for the MLX engines. \texttt{omlx} pages its prefix cache to SSD (fresh cache directory and server restart before each level) and ran without the memory sidecar. It streams roughly two tokens per SSE chunk, so its ITL cells are chunk-derived and not comparable. Runs of 2026-08-24/25.}
\label{tab:qwen38_27b}
\end{table}

\FloatBarrier

\clearpage
\subsection{Concurrent serving of 4-bit dense and MoE models}
\label{sec:appendix_large_model_serving}

A separate follow-up on September~1--3, 2026 compares Qwen3.8-27B
(dense) and Qwen3.6-35B-A3B (MoE, 3B active parameters) on the
64\,GB M5~Pro. It reuses the fixed agent prompts, with 100 requests
and three warmups at $c{=}1/2/4$. Each level starts a fresh server.
The plotted \texttt{omlx} configuration uses its default SSD-backed
prefix cache, reset between concurrency levels. \texttt{llama.cpp}
uses four server slots and GGUF \texttt{UD-Q4\_K\_M}; the MLX
engines share each model's \texttt{mlx-community} 4-bit conversion.
Quantization formats and weight footprints differ across these
configurations. The later builds and archived result files are
documented separately from the main campaign.\footnote{The paper's
\texttt{reproducibility/larger\_models\_2026\_09/} directory contains
the plotted runs and RAM-cache sensitivity results, with recorded
versions, settings, and checksums.}

\begin{figure}[H]
\centering
\includegraphics[width=\linewidth]{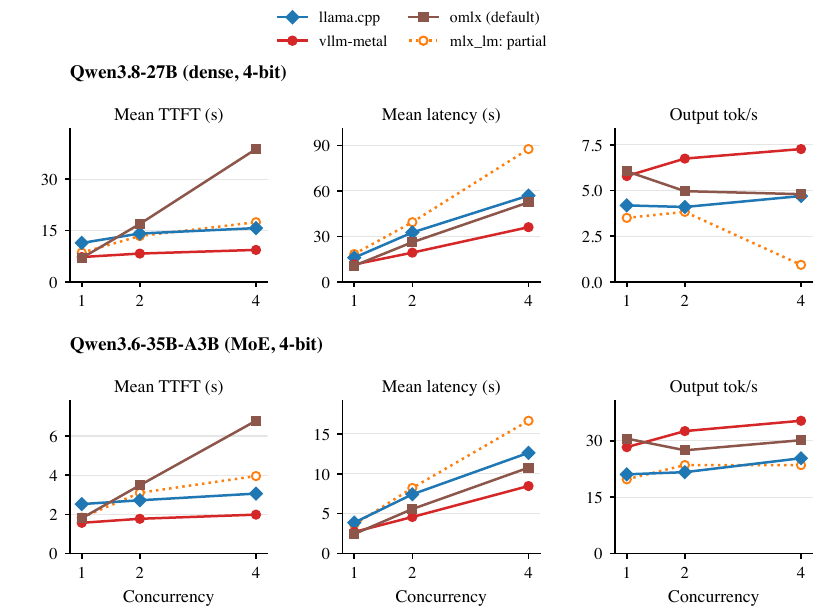}
\caption{\textbf{The MoE throughput gap narrows, while first-token
latency still separates the serving paths.} September 2026
follow-up, nominal 4-bit models on the M5~Pro. Latency panels show
means over completed requests; output throughput uses completed
output tokens over full sweep wall time. Hollow \texttt{mlx\_lm}
markers denote partial runs: 30/31/23 of 100 requests completed
at $c{=}1/2/4$ on 27B, and 38/40/38 on MoE. \texttt{llama.cpp}
completes 99/100 at every level; the remaining curves complete
100/100. Partial-run latency covers a different request subset.}
\label{fig:large_model_serving}
\end{figure}

Figure~\ref{fig:large_model_serving} supplements the throughput and TTFT
comparison in \S\ref{sec:results_large_models} with mean request latency.
On the dense 27B model, \texttt{vllm-metal} also has the lowest mean
request latency at $c{=}4$. Each configuration was measured once.
An archived RAM-cache sensitivity run narrows
\texttt{vllm-metal}'s MoE throughput advantage at $c{=}4$ from
$17\%$ to $4\%$.
\FloatBarrier
\clearpage

\section{DGX Spark reference track: full results}
\label{sec:appendix_dgxspark}

Hardware: NVIDIA DGX Spark (GB10 Grace-Blackwell, 128\,GB unified
LPDDR5x, Linux/CUDA). Engines: upstream \texttt{llama.cpp} (CUDA
build, GGUF weights), \texttt{vllm}, and \texttt{sglang}
(Safetensors weights), installed and served by the platform track
of the same harness (\texttt{--platform dgxspark}); identical
prompts, BF16 weights, and concurrency grid as the Apple track,
$n{=}100$ per level. Runs of 2026-07-03 to 2026-07-05.
Table~\ref{tab:dgxspark_qwen06} reports the primary model;
Table~\ref{tab:dgxspark_crossmodel} extends to the two further
model releases used in Appendix~\ref{sec:appendix_crossmodel}. The
memory and fidelity lenses do not extend to this track
(Appendix~\ref{sec:appendix_limitations}).

\begin{table}[h]
\centering
\small
\setlength{\tabcolsep}{4.5pt}
\begin{tabular}{@{}lrrrrrrr@{}}
\toprule
 & \multicolumn{3}{c}{\textbf{Output tok/s at $c$}} & & \multicolumn{3}{c}{\textbf{TTFT p50 at $c$}} \\
\cmidrule(lr){2-4} \cmidrule(l){6-8}
\textbf{Engine} & \textbf{1} & \textbf{8} & \textbf{16} & $\mathbf{c{=}16/c{=}1}$ & \textbf{1} & \textbf{8} & \textbf{16} \\
\midrule
\rowcolor{gray!16}[0pt][\tabcolsep]\multicolumn{8}{@{}l}{\textit{chat split}} \\
\texttt{llama.cpp} & 119.4 & 157.7 & 235.3 & 2.0$\times$ & 69\,ms & 138\,ms & 178\,ms \\
\texttt{vllm} & 103.3 & 633.3 & 796.0 & 7.7$\times$ & 27\,ms & 37\,ms & 52\,ms \\
\texttt{sglang} & 106.8 & 592.3 & 734.1 & 6.9$\times$ & 22\,ms & 33\,ms & 42\,ms \\
\midrule
\rowcolor{gray!16}[0pt][\tabcolsep]\multicolumn{8}{@{}l}{\textit{agent split}} \\
\texttt{llama.cpp} & 55.3 & 70.2 & 79.7 & 1.4$\times$ & 510\,ms & 1.18\,s & 1.90\,s \\
\texttt{vllm} & 80.2 & 294.1 & 342.7 & 4.3$\times$ & 78\,ms & 85\,ms & 134\,ms \\
\texttt{sglang} & 82.6 & 323.5 & 384.1 & 4.6$\times$ & 67\,ms & 58\,ms & 84\,ms \\
\bottomrule
\end{tabular}
\caption{\textbf{Upstream vLLM and SGLang scale $4.3$--$7.7\times$ with flat TTFT; llama.cpp plateaus on CUDA too.} DGX Spark reference track, Qwen3-0.6B BF16, $n{=}100$ per level. All cells completed 100/100 requests. Runs of 2026-07-03 to 2026-07-05.}
\label{tab:dgxspark_qwen06}
\end{table}

\begin{table}[h]
\centering
\small
\setlength{\tabcolsep}{4.5pt}
\begin{tabular}{@{}lrrrrrrr@{}}
\toprule
 & \multicolumn{3}{c}{\textbf{chat: tok/s at $c$}} & & \multicolumn{3}{c}{\textbf{agent: tok/s at $c$}} \\
\cmidrule(lr){2-4} \cmidrule(l){6-8}
\textbf{Engine} & \textbf{1} & \textbf{8} & \textbf{16} & & \textbf{1} & \textbf{8} & \textbf{16} \\
\midrule
\rowcolor{gray!16}[0pt][\tabcolsep]\multicolumn{8}{@{}l}{\textit{Qwen3.5-0.8B}} \\
\texttt{llama.cpp} & 106.5 & 162.1 & 376.1 & & 77.5 & 108.5 & 172.7 \\
\texttt{vllm} & 94.2 & 529.8 & 680.6 & & 73.4 & 223.9 & 248.5 \\
\texttt{sglang} & 95.0 & 580.2 & 935.3 & & 79.4 & 396.4 & 684.1 \\
\midrule
\rowcolor{gray!16}[0pt][\tabcolsep]\multicolumn{8}{@{}l}{\textit{Gemma-4-E4B-it}} \\
\texttt{llama.cpp} & 20.1 & 29.8 & 72.8 & & 18.1 & 24.6 & 51.7 \\
\texttt{vllm} & 17.2 & 136.6 & 238.5 & & 16.1$^{\dag}$ & 131.9$^{\dag}$ & 226.4$^{\dag}$ \\
\texttt{sglang} & 17.2 & 129.4 & 211.7 & & 16.0$^{\dag}$ & 125.8$^{\dag}$ & 201.9$^{\dag}$ \\
\bottomrule
\end{tabular}
\caption{\textbf{The scaling pattern holds across model releases.} DGX Spark reference track, two further models (BF16, $n{=}100$ per level). $^{\dag}$94/100 requests completed (six agent prompts rejected); throughput is over completed requests.}
\label{tab:dgxspark_crossmodel}
\end{table}

\section{Limitations}
\label{sec:appendix_limitations}

(1)~The main serving audit uses one Apple M5~Pro;
generation-specific kernel support may change rankings on
older Macs. The DGX Spark reference covers three engines on one
box, throughput and latency only, with no recurring maintenance cycle.
AMD's unified-memory APUs are not covered.
(2)~Qwen3-0.6B is the common model across all nine stacks; the two
newer core releases expose support gaps.
The larger-model studies (\S\ref{sec:results_large_models}, Appendix~\ref{sec:appendix_large_models})
and 35B multi-node comparison (\S\ref{sec:results_d3}) cover
partial rosters and do not establish full-roster transfer.
The September follow-up uses later builds and different
quantization formats, and evaluates throughput and latency only.
(3)~Fidelity rests on a single classification task (GMRID, 8
classes).
(4)~We do not run a designed repeat pass, so the campaign carries
no formal run-to-run variance estimate, and near-ties are reported
as ties. In three incidental telemetry repeats, 8 of 9 levels moved
by at most $5.6\%$; \texttt{omlx} at agent $c{=}16$ moved $20\%$.
The retained cache-isolation settings are documented in
Appendix~\ref{sec:framework_settings}.
(5)~We do not test models that require multiple machines to fit in
memory or scaling beyond two nodes.
(6)~The Qwen3-0.6B agent split lacks a memory trace for
\texttt{mlx\_lm}, after its telemetry repeat exceeded the
wall-clock cap (Appendix~\ref{sec:framework_settings}). It is
reported as unmeasured; the retained throughput and TTFT cells are
unaffected.



\end{document}